\documentclass[%
 reprint,
 amsmath,amssymb,
 aps,
]{revtex4-1}

\usepackage{graphicx}
\usepackage{dcolumn}
\usepackage{bm}

\def\beq{\begin{eqnarray}}
\def\eeq{\end{eqnarray}}
\def\nn{\nonumber\\}
\def\l({\left(}
\def\r){\right)}
\def\v{\upsilon}

\begin{document}

\preprint{YITP-17-12}

\title{Resolving the Ostrogradsky Ghost Problem for a Class of Scalar-tensor Theories}

\author{Chunshan Lin}
\email{chunshan.lin@yukawa.kyoto-u.ac.jp}
\affiliation{Yukawa Institute for Theoretical Physics, Kyoto University}

\date{\today}

\begin{abstract}
We perform a Hamiltonian analysis and prove that a large class of scalar-tensor theories whose Lagrangian is given by a generic function $F\l(K_{ab},N,R_{ab},\nabla_a,h^{ab},t\r)$ is free from the Ostrogradsky ghost, and thus these theories contain at most 3 degrees of freedom at classical level. 
\begin{description}
\item[PACS numbers]
\end{description}
\end{abstract}

\pacs{Valid PACS appear here}
\maketitle

{\bf Introduction}
Searching for self-consistent gravitation theories beyond general relativity is well motivated by both of theoretical and experimental aspects. Experimentally, we do not know gravity's properties and behaviour at the microscopic scale and cosmological scale. Theoretically, a unified quantum mechanical theory of gravitation has never been achieved.  At high energy scales, graviton loop diagrams diverge due to the negative dimension of gravitational coupling constant. In Horava-Lifshitz gravity \cite{Horava:2009uw}, an improved UV behaviour of the graviton propagator is obtained by assuming an anisotropic scaling  between space and time. The theory is constructed on a hyper-surface with temporal differmorphism invariance partially broken. The theory is essentially a scalar-tensor theory if we recover the general covariance. 

At another extreme scale, the dark energy, or cosmological constant problem, is a longstanding problem in theoretical physics, and has been taken more and more seriously since the discovery of the accelerated expansion of our universe. One promising approach to the late time cosmic acceleration enigma is to introduce new degrees of freedom to the gravitational field. These new degrees of freedom might speed up the universe. There are several interesting theories of this kind in the literature, for instance quintessence\cite{Ratra:1987rm}\cite{Caldwell:1997ii}, k-essence\cite{ArmendarizPicon:2000dh} and ghost condensate\cite{ArkaniHamed:2003uy}.

In the physics of very early universe, it is believed that our universe underwent an exponential expansion phase which  we call inflation. Conventionally, inflation is driven by the vacuum energy stored in a scalar field slowly rolling on the flat potential. Inflation thus provides a natural platform for the extension of general relativity since it is equivalent to a scalar-tensor theory. 
A unified framework describing the low energy properties of scalar and tensor degrees is given in the language of effective field theory in Ref. \cite{Cheung:2007st}. 

However, with higher derivatives included,  the theory generally suffers from Ostrogradsky ghost instability \cite{Ostrogradsky}\cite{Woodard:2006nt}.  It is therefore intriguing to ask for the most general scalar-tensor theory which is free from the Ostrogradsky ghost. The attempts in this direction can be traced back to Horndeski's work in 1974 \cite{Horndeski} and it was rediscovered recently in the context of the so-called Galileon theory \cite{Nicolis:2008in,Deffayet:2009wt,Deffayet:2011gz}. It turns out that Horndeski's theory isn't the most general scalar-tensor theory with only three degrees of freedom. Recently, several examples beyond Horndeski theory have been spotted in the literatures which are free from the Ostrogradsky ghost \cite{Zumalacarregui:2013pma,Gleyzes:2014dya,Gao:2014soa,Lin:2014jga,Gao:2014fra,Langlois:2015cwa,BenAchour:2016fzp}.

In this letter, we provide a more general framework to prove a ghost freeness theorem. We work in the $3+1$ formalism and write down all possible geometrical quantities projected on the constant time hyper-surface. This is a large class of theories. By means of Dirac's Hamiltonian analysis method \cite{dirac1,dirac2,dirac3}, we prove that this large class of theories is free from the Ostrogradsky ghost and thus only contains $at~most$ three degrees of freedom. 

{\bf Hamiltonian Analysis}
In our scalar-tensor theory, the temporal diffeomorphism invariance is spontaneously broken due to the time-like vacuum expectation value of the scalar field. We define the constant time hyper-surface $\mathcal{M}$, and all vectors and tensors constructed out of the higher derivatives of scalar field in $3+1$ dimensional space-time are projected onto the constant time hyper-surface $\mathcal{M}$. The residual gauge symmetry on this hyper-surface is $x^i\to x^i+\xi^i(t,\bf{x})$. In the unitary gauge, the action is written in the following general form
\beq\label{genst1}
S=\int dtd^3x\sqrt{h}F\l(K_{ab},N,R_{ab},\nabla_a,h^{ab},t\r),
\eeq
where $K_{ab}$ and $R_{ab}$ are extrinsic curvature and Ricci tensor of the 3 dimensional hyper-surface respectively, $\nabla_{a}$ is the covariant derivative compatible with the induced metric $h_{ab}$ (we will  use $\nabla_a\mathcal{O}$ and $\mathcal{O}_{|a}$ interchangeably if no confusion is caused). All terms in the theory must be contracted to a scalar via the induced metric $h^{ab}$. We will consider a general scalar function $F$ and, since 
we are interested in physical models, we shall assume that 
any appropriate smoothness conditions are satisfied.

In what follows, we are not going to specify the detailed form of the function $F$, and try to prove a no-ghost theorem in the unified framework. Nevertheless, some obviously non-dynamical Lagrangians, i.e. the ones without the terms non-linear in extrinsic curvature $K_{ab}$, should be excluded from the outset. Noted that we have operator $\nabla_a$ included in the action,  it implies that this theory includes higher than second derivatives on the scalar field.  It seems very difficult to work out the Hamiltonian in terms of a local function. To  proceed with the Hamiltonian analysis by means of Dirac's method, we have to introduce an auxiliary tensor field $Q_{ab}$, and the equivalent theory is written as 
\beq
S=\int dtd^3x\sqrt{h}\left[F\l(Q_{ab},N,R_{ab},\nabla_a,h^{ab},t\r)\right.\nn
\left.+N\v^{ab}\l(Q_{ab}-K_{ab}\r)\right].
\eeq 
The last term in the above equation is a constraint. After solving this constraint, we recover our original theory eq. (\ref{genst1}).  The conjugate momenta are calculated as 
\beq\label{momenta}
\pi^{ab}&=&\frac{\partial\mathcal{L}}{\partial\dot{h}_{ab}}=-\frac{1}{2}\sqrt{h}\v^{ab},\qquad \pi_N=\frac{\partial\mathcal{L}}{\partial\dot{N}}=0,\nn
\pi_i&=&\frac{\partial\mathcal{L}}{\partial\dot{N}^{i}}=0,
\qquad\qquad~~~ P^{ab}=\frac{\partial\mathcal{L}}{\partial\dot{Q}_{ab}}=0,\nn U_{ab}&=&\frac{\partial\mathcal{L}}{\partial\dot{\upsilon}^{ab}}=0.
\eeq
The Hamiltonian reads
\beq
H&=&\int d^3x\left[ \pi^{ab}\dot{h}_{ab}-\mathcal{L}+\lambda_N\pi_N+\lambda^i\pi_i
+\chi_{ab}P^{ab}\right.\nn
&&\qquad\qquad\qquad\qquad\qquad~\left.+\varphi^{ab}U_{ab}+\lambda_{ab}\Psi^{ab}\right]\nn
&=&\int d^3x\left[ \mathcal{H}+N^i\mathcal{H}_i+\lambda_N\pi_N+\lambda^i\pi_i+\chi_{ab}P^{ab}\right.\nn
&&~\qquad\qquad\qquad\qquad\qquad\left.+\varphi^{ab}U_{ab}+\lambda_{ab}\Psi^{ab}\right],
\eeq
where 
\beq
\mathcal{H}&=&-\sqrt{h}\left[F\l(Q_{ab},N,R_{ab},\nabla_a,h^{ab},t\r)+N\v^{ab}Q_{ab}\right],\nn
\mathcal{H}_i&=&\sqrt{h}\nabla_j\v_{~i}^{j},\nn
\Psi&\equiv&\pi^{ab}+\frac{1}{2}\sqrt{h}\v^{ab},
\eeq
and we have used $\pi^{ab}=-\frac{1}{2}\sqrt{h}\v^{ab}$ in the eq. (\ref{momenta}) to eliminate velocity $\dot{h}_{ab}$ in the Hamiltonian. There are five primary constraints
\beq
\pi_N&\approx&0,\qquad\pi_i\approx0,\qquad P^{ab}\approx0,\nn
\qquad U_{ab}&\approx&0,\qquad\Psi^{ab}\approx0.
\eeq
To be consistent, these five primary constraints must be independent of time. The consistency conditions give us three secondary constraints,
\beq\label{secons}
\frac{d\pi_N}{dt}&=&\{\pi_N,H\}
\approx\sqrt{h}\l(\frac{\partial F}{\partial N}+\v^{ab}Q_{ab}\r)\equiv \mathcal{C}\approx 0,\nn
\frac{d\pi_i}{dt}&=&\{\pi_i,H\}=-\mathcal{H}_i=-\sqrt{h}\nabla_j\v^j_{~i}\approx0,\nn
\frac{dP^{ab}}{dt}&=&\{P^{ab},H\}
\approx\sqrt{h}\left(\frac{\partial F}{\partial Q_{ab}}+N\v^{ab}\right)\nn
&&\qquad\qquad\equiv\Phi^{ab}\approx0,
\eeq
where $\{...\}$ denotes the Poisson bracket.
To avoid the possible confusion, the partial differentiation should be understood in terms of integral form
\begin{equation}
\frac{\partial F}{\partial q}\equiv \int \frac{\partial F}{\partial q},\qquad \frac{\partial^2 F}{\partial q_1\partial q_2}\equiv \int \frac{\partial}{\partial q_2}\int \frac{\partial F}{\partial q_1}.
\end{equation}
 The algebra closes here because the consistency conditions for the rest of two primary constraints $U_{ab}\approx0,~\Psi^{ab}\approx0$ and  three new secondary constraints in eq. (\ref{secons}) only fix the coefficients in front of them. Let us collect all of primary and secondary constraints in the total Hamiltonian and treat all of them on the same footing, 
\beq
H_{\text{tot}}=\int d^3x\left[ \mathcal{H}+N^i\mathcal{H}_i+\lambda_N\pi_N+\lambda^i\pi_i+\chi_{ab}P^{ab}\right.\nn
\left.+\varphi^{ab}U_{ab}+\lambda_{ab}\Psi^{ab}+\phi_{ab}\Phi^{ab}+\lambda_c\mathcal{C}\right].~
\eeq
In passing, we note that if the symmetry of our theory is enhanced, i.e. invariant under all space-time diffeomorphisms $x^\mu\to x^\mu+\xi^\mu(t,\textbf{x})$, the constraints $\mathcal{C}\approx0,\pi_N\approx0$ are first class, and Hamiltonian constraint $\mathcal{C}\approx0$ plays the role of temporal coordinate transformation generator. This is the theory of Einstein gravity according to Lovelock theorem \cite{lovelock1,lovelock2}. In our current work, we are not going to consider the case with enhanced symmetry. 

It is easy to check that $\pi_i\approx0$ is first class and it eliminates the conjugate canonical pair $\left(N^i,\pi_i\right)$. On the other hand, we have $\mathcal{C}\approx0,\pi_N\approx0,\Phi^{ab}\approx0,P^{ab}\approx0,\Psi^{ab}\approx0$ and $U_{ab}\approx0$ which are all second class, eliminating the canonical variable pairs $\left(N,\pi_N\right)$, $\l(Q_{ab},P^{ab}\r)$ and $\l(\v^{ab},U_{ab}\r)$ respectively. The only conjugate canonical pair which remains as would-be dynamical degrees is the $\l(h_{ab},\pi^{ab}\r)$. The number of dynamical degrees of freedom crucially depends on the property of three momentum constraints $\mathcal{H}_i\approx0$. We have checked that their Poisson brackets with constraints  $\mathcal{C}\approx0,\Phi^{ab}\approx0,\Psi^{ab}\approx0$ and $U_{ab}\approx0$   do not vanish weakly. But the extended momentum constraints
\beq\label{emc}
\bar{\mathcal{H}}_{\text{E}}[N^i]&\equiv&\int N^i\mathcal{H}_i+2N_{i|j}\Psi^{ij}+A_{ij}P^{ij}+B^{ij}U_{ij}+C\pi_N\nn
&\approx&0
\eeq 
could be first class, given the proper coefficients $A_{ij}, B^{ij}, \text{and}~ C$. 
We can check that all of the following Poisson brackets weakly vanish,
\beq
&&\{\bar{\mathcal{H}}_{E}[N^i],\bar{\mathcal{H}}_{E}[f^i]\}\approx0,\qquad\{\bar{\mathcal{H}}_{E}[N^i], \bar{P}[\chi_{ab}]\}\approx0,\nn
&&\{\bar{\mathcal{H}}_{E}[N^i], \bar{U}[\varphi^{ab}]\}\approx0,\qquad
\{\bar{\mathcal{H}}_{E}[N^i], \bar{\pi}_N[\lambda_N]\}\approx0,\nn
&&\{\bar{\mathcal{H}}_{E}[N^i], \bar{\pi}_i[\lambda^i]\}\approx0,
\eeq
where $\bar{\mathcal{O}}[\lambda]\equiv \int  d^3x\lambda \mathcal{O}$ with all indices omitted.
There are only three non-vanishing Poisson brackets
\begin{widetext}\beq
&&\{\bar{\mathcal{H}}_{E}[N^i], \bar\Phi[\phi_{ab}]\}\approx
-\int \sqrt{h}\left[2N_{i|j}\frac{\partial^2F}{\partial Q_{ab}\partial h_{ij}}+\frac{\partial^2F}{\partial Q_{ab}\partial Q_{ij}}A_{ij}+NB^{ab}+C\l(\frac{\partial^2 F}{\partial N\partial Q_{ab}}+\v^{ab}\r)\right]\phi_{ab},\label{Ht1}\\
&&\{\bar{\mathcal{H}}_{E}[N^i], \bar\Psi[\lambda_{ab}]\}\approx -\frac{1}{2}\int\sqrt{h}\left[-N^a\v^{ij}_{~~|a}+\v^{ai}N^j_{~|a}+\v^{aj}N^i_{~|a}+B^{ij}\right]\lambda_{ij},\label{Ht2}\\
&&\{\bar{\mathcal{H}}_{E}[N^i], \bar{\mathcal{C}}[\lambda_c]\}\approx -\int \sqrt{h}\left[2N_{i|j}\frac{\partial^2F}{\partial N\partial h_{ij}}+A_{ij}\l(\frac{\partial^2F}{\partial N\partial Q_{ij}}+v^{ij}\r)+B^{ij}Q_{ij}+\frac{\partial^2F}{\partial N^2}C\right]\lambda_c.\label{Ht3}
\eeq
\end{widetext}
We can always find the proper $A_{ij}$, $B_{ij}$ and $C$ so that the Poisson brackets of eq. (\ref{Ht1})(\ref{Ht2}) and (\ref{Ht3}) also vanish. One may worry that the above equations are degenerate in the case that Lagrangian is linear in the lapse $N$. However, the residual gauge symmetries on the hyper-surface always enforce the third Poisson bracket to vanish weakly, given the proper $A_{ij}$ and $B^{ij}$ so that the first two Poisson brackets vanish weakly.  Therefore, we come up with a much easier method to compute $A_{ij}$, $B^{ij}$ and $C$. The residual diffeomorphisms $x^i\to x^i+\xi^i(t,\bf{x})$ invariance  on the hyper-surface $\mathcal{M}$  implies that there must be three first class momentum constraints playing the role of coordinate transformation generators. The scalar and tensor must transform accordingly,
\beq
\{h_{ab},\bar{\mathcal{H}}_{E}[\xi^i]\}&=&\xi_{a|b}+\xi_{b|a},\\
\{Q_{ab},\bar{\mathcal{H}}_{E}[\xi^i]\}&=&\xi^cQ_{ab|c}+\xi^c_{~|a}Q_{cb}+\xi^c
_{~|b}Q_{ca},\\
\{\v^{ab},\bar{\mathcal{H}}_{E}[\xi^i]\}&=&\xi^c\v^{ab}_{~~|c}-\xi^a_{~|c}\v^{cb}-\xi^b_{~|c}\v^{ca},\\
\{N,\bar{\mathcal{H}}_{E}[\xi^i]\}&=&\xi^i\nabla_iN.
\eeq
The $A_{ij}$, $B_{ij}$ and $C$ can be worked out by solving the above equations, and the extended momentum constraints read
\beq\label{emc2}
\mathcal{H}_{E,i}&=&-2\sqrt{h}\left(\frac{\pi^{j}_{~i}}{\sqrt{h}}\right)_{|j}+Q_{ab|i}P^{ab}-2\left(P^{ab}Q_{ia}\r)_{|b}\nn
&&~\qquad+\v^{ab}_{~~|i}U_{ab}+2\l(\v^{aj}U_{ij}\r)_{|a}+\pi_N\partial_iN.
\eeq
Given the extended momentum constraints of eq. (\ref{emc2}), their Poisson brackets with all constraints vanish trivially and weakly  due to the fact that we have constructed our theory on the constant time hyper-surface $\mathcal{M}$, and thus the Hamiltonian as well as all constraints are manifestly invariant under the coordinate transformations generated by the  extended momentum constraints. Explicitly, we have 
\beq
&&\{\bar\Phi[\phi_{ab}], \bar{\mathcal{H}}_{E}[\xi^i] \}\nn
&\approx& \int  \frac{\partial\bar\Phi }{\partial h_{ij}}\delta_{\xi}h_{ij}+ \frac{\partial\bar\Phi }{\partial Q_{ij}}\delta_{\xi}Q_{ij}+ \frac{\partial\bar\Phi }{\partial \v^{ij}}\delta_{\xi}\v^{ij}+ \frac{\partial\bar\Phi }{\partial N}\delta_{\xi}N\nn
&=&\int \delta_\xi\bar\Phi[\phi_{ab}]\approx0,
\eeq
where $\delta_{\xi}\mathcal{O}$ denotes the variation of $\mathcal{O}$ under the differmorphisms $x^i\to x^i+\xi^i(t,\bf{x})$. Similarly, we have $\{\bar\Psi[\lambda_{ab}],\bar{\mathcal{H}}_{E}[\xi^i]\}\approx \int \delta_\xi\bar\Psi[\lambda_{ab}]\approx0$ and $\{ \bar{\mathcal{C}}[\lambda_c], \bar{\mathcal{H}}_{E}[\xi^i]\}\approx \int \delta_\xi \bar{\mathcal{C}}[\lambda_c]\approx 0$. 

Therefore, we have 3 extended momentum constraints which are first class  and they kill 6 degrees in the phase space of conjugate pair $\l(h_{ab},\pi^{ab}\r)$. We conclude that there are $at~most$  three dynamical degrees of freedom in our theory. We have proved that a very large class of theories beyond Horndeski is actually free from the Ostrogradsky ghost.

Let's end this section with several remarks. Firstly, we note that our no-ghost theorem still holds if we include the 3 dimensional covariant Levi-Civita tensor $\varepsilon^{abc}$ as the one of arguments of  function $F$ in our theory, at the cost of the parity violation.  The generalisation to this case is trivial, and thus we are not going to provide a new proof. 

Secondly, we have to reiterate that the number of degrees of freedom in our theory is $at~most$ three. The phase space of conjugate pair $\l(h_{ab},\pi^{ab}\r)$ is further diminished by one degree if $\pi_N\approx0$ upgrade to first class (even in the phase $t\to t+\xi^0(t,\textbf{x})$ symmetry is broken).  This is the case when the function $F$ is linear in lapse $N$ and meanwhile free from mixed derivative term such as $\nabla K\nabla K$. The theories with diminished phase space is beyond the scope of our current work, and will be discussed in our future work \cite{futurework}.

Thirdly, our analysis does not include the free shift $N^i$ in the action (except for those ones initially included in $K_{ij}$), because under the  spatial diffeomorphisms  $x^i\to x^i+\xi^i(t,\bf{x})$, the shift does not transform as a vector on the hyper-surface $\mathcal{M}$. The shift in the action generally gives rise to additional degrees of freedom.  Neverthless, if the shift appears in the action in terms of the following combinations  \cite{Coates:2016zvg}
\beq
r_{ij}&\equiv&\frac{1}{N}\l(\dot{R}_{ij}-N^kR_{ij|k}-R_{ik}N^k_{~|j}-R_{ik}N^k_{~|j}-R_{jk}N^k_{~|i}\r),\nn
\mathcal{A}_i&\equiv&\frac{1}{N}\l(\dot{a}_i-N^ja_{i|j}-N^j_{~|i}a_j\r),
\eeq
where $a_i\equiv \partial_i\ln N$, the residual diffeomorphisms invariance $x^i\to x^i+\xi^i(t,\bf{x})$ is still preserved and thus the extended momentum constraints, with more counter terms included probably, are still first class. However, after doing some simple maths, we find that $r_{ij}$ is equivalent to the terms of covariant derivatives of lapse $N$ and extrinsic curvature $K_{ab}$, i.e. $\nabla_{..}\nabla_{..}\l(NK^{..}\r)$, and thus it does not give us any new terms. On the other hand, if we have $\mathcal{A}_i$ in the action, the conjugate pair $\l(N,\pi_N\r)$ becomes dynamical, and we have two scalar degrees in the theory.  One of them is the Ostrogradsky ghost, and the Hamiltonian is unbounded from below. The instability was perturbatively demonstrated in Ref. \cite{Coates:2016zvg} in the context of the extended theory of Horava-Lifshitz gravity. Thus we should exclude the terms with $ \mathcal{A}_i$ in our no-ghost theorem.

{\bf Conclusion and Discussion}
In this letter,  we performed a Hamiltonian analysis by means of Dirac's approach for a large class of scalar-tensor theories whose Lgrangian is given by a generic function of $F\l(K_{ab},N,R_{ab},\nabla_a,h^{ab},t\r)$. We find three first class momentum constraints eliminating 6 degrees on the phase space. They are the coordinate transformation generators on the constant time hyper-surface $\mathcal{M}$. We have carefully counted the number of degrees of freedom in the theory, and we found that this class of theories contains $at~most$ three degrees of freedom. We conclude that this class of theories is free from the Ostrogradsky ghost.

With a large class of Ostrogradsky ghost free theories at hand, it is tempting to ask how to alleviate or even solve these open problems in modern physics of gravitation, for instance, quantum gravity, the cosmological constant problem,  the origin of hot big bang and so on. We hope to come back to these issues in the future. 
\\

{\bf Acknowledgments}
We would like to thank S. Mukohyama,  and  R. Saito  for useful discussions, and thank M. Sasaki and M. C. Werner for useful discussions and the suggestions on improving the manuscript. This work is supported by JSPS  fellowship for overseas researchers.


\begin{thebibliography}{99}
\bibitem{Horava:2009uw} 
  P.~Horava,
  Phys.\ Rev.\ D {\bf 79}, 084008 (2009)
  [arXiv:0901.3775 [hep-th]].
  
  \bibitem{Ratra:1987rm} 
  B.~Ratra and P.~J.~E.~Peebles,
  Phys.\ Rev.\ D {\bf 37}, 3406 (1988).
\bibitem{Caldwell:1997ii} 
  R.~R.~Caldwell, R.~Dave and P.~J.~Steinhardt,
  Phys.\ Rev.\ Lett.\  {\bf 80}, 1582 (1998)
  doi:10.1103/PhysRevLett.80.1582
  [astro-ph/9708069].
  
\bibitem{ArmendarizPicon:2000dh} 
  C.~Armendariz-Picon, V.~F.~Mukhanov and P.~J.~Steinhardt,
  Phys.\ Rev.\ Lett.\  {\bf 85}, 4438 (2000)
  [astro-ph/0004134].
  

  
\bibitem{ArkaniHamed:2003uy} 
  N.~Arkani-Hamed, H.~C.~Cheng, M.~A.~Luty and S.~Mukohyama,
  JHEP {\bf 0405}, 074 (2004)
  [hep-th/0312099].

  
\bibitem{Cheung:2007st} 
  C.~Cheung, P.~Creminelli, A.~L.~Fitzpatrick, J.~Kaplan and L.~Senatore,
  JHEP {\bf 0803}, 014 (2008)
  [arXiv:0709.0293 [hep-th]].
\bibitem{Ostrogradsky}
M. Ostrogradskyy: Mem. Ac. St. Petersbourg VI 4, 385 (1850).
\bibitem{Woodard:2006nt} 
  R.~P.~Woodard,
  Lect.\ Notes Phys.\  {\bf 720}, 403 (2007)
  [astro-ph/0601672].

\bibitem{Horndeski}
 G. W. Horndeski, Int. J. Theor. Phys. 10, 363 (1974).
\bibitem{Nicolis:2008in} 
  A.~Nicolis, R.~Rattazzi and E.~Trincherini,
  Phys.\ Rev.\ D {\bf 79}, 064036 (2009)
  [arXiv:0811.2197 [hep-th]].
\bibitem{Deffayet:2009wt} 
  C.~Deffayet, G.~Esposito-Farese and A.~Vikman,
  Phys.\ Rev.\ D {\bf 79}, 084003 (2009)
  [arXiv:0901.1314 [hep-th]].
  
\bibitem{Deffayet:2011gz} 
  C.~Deffayet, X.~Gao, D.~A.~Steer and G.~Zahariade,
  Phys.\ Rev.\ D {\bf 84}, 064039 (2011)
  [arXiv:1103.3260 [hep-th]].




  \bibitem{horndeski}
G. W. Horndeski, Int. J. Theor. Phys. 10, 363 (1974).
  
  
\bibitem{Zumalacarregui:2013pma} 
  M.~Zumalac\'arregui and J.~Garc\'ia-Bellido,
  Phys.\ Rev.\ D {\bf 89}, 064046 (2014)
  [arXiv:1308.4685 [gr-qc]].
  
  
\bibitem{Gleyzes:2014dya} 
  J.~Gleyzes, D.~Langlois, F.~Piazza and F.~Vernizzi,
  Phys.\ Rev.\ Lett.\  {\bf 114}, no. 21, 211101 (2015)
  [arXiv:1404.6495 [hep-th]].
\bibitem{Gao:2014soa} 
  X.~Gao,
  Phys.\ Rev.\ D {\bf 90}, 081501 (2014)
  [arXiv:1406.0822 [gr-qc]].
\bibitem{Gao:2014fra} 
  X.~Gao,
  Phys.\ Rev.\ D {\bf 90}, 104033 (2014)
  [arXiv:1409.6708 [gr-qc]].
\bibitem{Lin:2014jga} 
  C.~Lin, S.~Mukohyama, R.~Namba and R.~Saitou,
  JCAP {\bf 1410}, no. 10, 071 (2014)
  [arXiv:1408.0670 [hep-th]].
\bibitem{Langlois:2015cwa} 
  D.~Langlois and K.~Noui,
  JCAP {\bf 1602}, no. 02, 034 (2016)
  [arXiv:1510.06930 [gr-qc]].

\bibitem{BenAchour:2016fzp} 
  J.~Ben Achour, M.~Crisostomi, K.~Koyama, D.~Langlois, K.~Noui and G.~Tasinato,
  JHEP {\bf 1612}, 100 (2016)
  [arXiv:1608.08135 [hep-th]].

\bibitem{dirac1}
 P. A. Dirac, Can. J. Math. 2 129 (1950).
 
\bibitem{dirac2} 
 P. A. Dirac, Proc. Roy. Soc. London, ser. A, 246, 326 (1958).

\bibitem{dirac3}
P. A. M. Dirac, Lectures on Quantum Mechanics (Yeshiva University, New York 1964).

\bibitem{lovelock1}
D. Lovelock, J. Math. Phys. 12, 498 (1971).

\bibitem{lovelock2}
D. Lovelock, J. Math. Phys. 13, 874 (1972).

\bibitem{futurework}
C. Lin, et al., in progress.

\bibitem{Coates:2016zvg} 
  A.~Coates, M.~Colombo, A.~E.~Gumrukcuoglu and T.~P.~Sotiriou,
  Phys.\ Rev.\ D {\bf 94}, no. 8, 084014 (2016)
  [arXiv:1604.04215 [hep-th]].
  
\end{thebibliography}
\end{document}